\documentclass{article}

\usepackage{eusipco}
\usepackage{epsfig}
\title{Comparison of the estimation of the degree of polarization from four or two intensity images degraded by speckle noise}

\name {Muriel Roche, Philippe R\'efr\'egier}

\address{ Physics and Image Processing Group -
Fresnel Institute \\
Unit\'e Mixte de Recherche 6133 -
Ecole G\'en\'eraliste d'Ing\'enieurs de Marseille\\
Domaine Universitaire de Saint-J\'er\^ome - 13397 Marseille cedex
20 France\\
phone: + (33 )  4 91 28 80 86, fax: + (33) 4 91 28 82 01, email: muriel.roche@fresnel.fr \\
web: www..fresnel.fr/phyti/}

\begin{document}

\maketitle

%% PLACE YOUR ABSTRACT HERE
\begin{abstract}
Active polarimetric imagery is a powerful tool for accessing the information present in a scene. Indeed, the polarimetric images obtained can reveal polarizing properties of the objects that are not avalaible using conventional imaging systems \cite{wol94,ref06a}. However, when coherent light is used to illuminate the scene, the images are degraded by speckle noise.
The polarization properties of a scene are characterized by the degree of polarization. In standard polarimetric imagery system, four intensity images are needed to estimate this degree \cite{bro98}. If we assume the uncorrelation of the measurements, this number can be decreased to two images using the Orthogonal State Contrast Image (OSCI) \cite{gou01}. However, this approach appears too restrictive in some cases. We thus propose in this paper a new statistical parametric method to estimate the degree of polarization assuming correlated measurements  with only two intensity images. The estimators obtained from four images, from the OSCI and from the proposed method, are compared using simulated polarimetric data degraded by speckle noise.
\end{abstract}

%% AND NOW START WITH YOUR PAPER CONTENT

%%%%%%%%%%%%%%%%%%%%%%%%%%%%%%%%%%%%%%%%%%%%%%%%%%%%%%%%%%%%%%%%%%
\section{Introduction}
%%%%%%%%%%%%%%%%%%%%%%%%%%%%%%%%%%%%%%%%%%%%%%%%%%%%%%%%%%%%%%%%%%

Polarimetric imagery consists in forming an image of the state of polarization of the light backscattered by a scene. We consider in this paper that the scene is artificially illuminated with coherent light (laser). For example, this illumination is used in active imagery in order to combine night vision capability and to improve image resolution for a given aperture size. In practice, using a coherent illumination produces speckle noise that deteriorate the image \cite{goo85}. However, the backscattered light gives information about the capability of the scene to polarize or depolarize the emitted light and thus allows one to determine the medium that compose the scene. These information can be described by a scalar parameter: the degree of polarization of light. This quantity is obtained in standard configurations of polarimetric systems using four pair of angular rotations of both a compensator and a polarizer. Four transmittance are thus recorded \cite{bro98} that lead to the estimation of the degree of polarization. However, this system is complex and it is interesting to develop methods to estimate the degree of polarization that could reduce the number of images to register. In \cite{gou01}, the authors proposed to estimate the degree of polarization with only two intensity images, however this method relies on the assumption that the measurements of the two components are uncorrelated which can be in some cases a too restrictive hypothesis. This paper extends the work of \cite{gou01} by taking into account the correlation of the different components.\\
 Let us first introduce the context of the study.\\

\section{Background}
The electric field of the light at a point of coordinates { \bf r} (vector of 3 component) in a 3D space and at time $t$ can be written, if we assume the light to propagate in a homogeneous and isotropic medium, as
\begin{equation}
 {\bf E}({\bf r},t) = \left[ A^{X}({\bf r},t) {\bf e_x} + A^{Y}({ \bf r},t) {\bf e_y}\right]. e^{- i 2 \pi \nu t}
\end{equation}
where $\nu$ is the central frequency of the field and $\bf{e_x}, \bf{e_y}$ are unitary orthogonal vectors (in the following bold letters represent vectors).\\
The terms $A^{X}({\bf r},t)$ and $ A^{Y}({\bf r},t)$ are complex and define the random vector called Jones vector
\begin{equation}
\bf{A} = \left[\begin{array}{l}
   A^{X}({\bf r},t) \\
   A^{Y}({\bf r},t)
\end{array}\right].
\end{equation}

The state of polarization of light corresponds to the properties of ${\bf{E}}({\bf r},t)$ at a particular point of space. It can be described by the covariance matrix $\Gamma$

\begin{equation}
\Gamma = \left[\begin{array}{ll}
   <A^{X*}A^{X}> & <A^{X*}_{\lambda}A^{Y}> \\
   <A^{Y*} A^{X}>&<A^{Y*}_{\lambda}A^{Y}>  
   \end{array}\right]
\end{equation}
where $<.>$ and $.^*$ define respectively the statistical average and the complex conjugate. For sake of brevity, the following notations for $\Gamma$ are introduced
\begin{equation}\label{cov}
\Gamma = \left[\begin{array}{ll}
   a_1 & a_2 \\
   a_2^* & a_4 \\
   \end{array}\right] = \frac{1}{c_1c_4-|c_2|^2} \left[\begin{array}{ll}
   c_4 & -c_2 \\
   -c_2^* & c_1 \\
   \end{array}\right]; 
\end{equation}
\begin{equation}
\hspace{0.6cm} \Gamma^{-1} = \left[\begin{array}{ll}
   c_1 & c_2  \\
   c_2^* & c_4 \\
   \end{array}\right] 
\end{equation}
Let us note that this matrix can be diagonalized since it is hermitic. \\
 In the case of coherent light, the electric field is represented by the complex Jones vector $\bf{A}$ which follows a Gaussian circular law \cite{goo85}
\begin{equation}
 p_{\Gamma}({\bf{A}}) = \frac{1}{\pi^2 det(\Gamma)} e^{-{\bf{A}}^{\dagger}  \Gamma^{-1}{\bf{A}}} 
\end{equation}
where ${\bf A}^{\dagger}$ stands for the adjoint of the vector ${\bf{A}}$.\\

The degree of polarization is defined by  \cite{goo85}

\begin{equation}
P^2= \frac{\mu_1 - \mu_2}{\mu_1 + \mu_2}
\end{equation}
where $\mu_1$ and $\mu_2$ are the eigenvalues of $\Gamma$ ($\mu_1 \geq \mu_2 \geq 0$).\\
The degree of polarization is a scalar parameter that characterizes the state of polarization of the light: if $P = 0$, the light is said to be totally depolarized, and if  $P = 1$, the light is said to be totally polarized. In the intermediate cases, the light is partially polarized.\\
With the notations introduced in (\ref{cov}) one can show that \cite{goo85}
\begin{equation}\label{deg_pol}
P^2= 1 - \frac{4 (a_1 a_4 - |a_2|^2)}{(a_1 + a_4)^2}.
\end{equation}
The knowledge of this quantity allows one to study the way the illuminated scene 
polarized or depolarized the emitted light. This degree gives information about the nature of the medium in the scene. In the standard configuration, four measurements are needed to estimate it. In the case of uncorrelated measurements, two intensity images gives a good estimation of this degree using the OSCI \cite{gou01}. However, in some cases the uncorrelation of the measurements may be not valid.\\
In this paper, an original estimation method that both uses a pair of images and that accounts for correlation in the components is proposed. We recall in the following the method proposed in \cite{gou01} and we extend it to correlated measurements. We then compare them through statistical measures using simulated data. The results are also presented when the standard estimation of $P^2$ is used (four images) as this case is expected to give the best results. Finally we conclude and give the perspectives of this work.
%%%%%%%%%%%%%%%%%%%%%%%%%%%%%%%%%%%%%%%%%%%%%%%%%%%%%%%%%%%%%%%%%%
\section{The OSCI}
%%%%%%%%%%%%%%%%%%%%%%%%%%%%%%%%%%%%%%%%%%%%%%%%%%%%%%%%%%%%%%%%%%
In the case of two uncorrelated images, the Orthogonal State Contrast Image (OSCI) is determined from \cite{gou01}, \cite{bre99}, \cite{gou04b}
\begin{equation}
\rho(i) = \frac{I_1(i) - I_2(i)}{I_1(i) + I_2(i)}
\end{equation}
where $I_1(i)$ and $I_2(i)$ are intensity measurements at the pixel site $(i)$, assuming a lexicographic order for the pixels.\\
These two images are obtained with simple polarimetric systems. First, the scene is illuminated by coherent light with single elliptical polarization state. Then the backscattered light is analysed in the polarization state parallel (which leads to $I_1(i) = |A_X(i)|^2)$ and orthogonal (which leads to $I_2(i) = |A_Y(i)|^2) $ to the incident one. \\
The OSCI is an estimation of $P^2$ in each pixel provided that the materials of the scene  modify the degree of polarization of incident light without modifying its principal polarized state (\footnote{State represented by the eigenvector associated to eigenvalue $\mu_1$.}). Let us recall that this kind of material is called pure depolarizer. For such materials, the covariance matrix $\Gamma$ is diagonal and, since the diagonal terms represent the intensity images, the OSCI gives an estimation of $P^2$ with 
\begin{equation}\label{posci}
{\hat P}_{OSCI}^2(i) = \left(\frac{I_1(i) - I_2(i)}{I_1(i) + I_2(i)}\right)^2 = \eta^2(i).
\end{equation}
In the case of non pure depolarizer objects (i.e. $\Gamma$ is non diagonal), the OSCI still reveals interesting contrast image but no more defines $P^2$. This leads us to a new method that considers the cases of correlated  measurements. This is the object of the following part.

%%%%%%%%%%%%%%%%%%%%%%%%%%%%%%%%%%%%%%%%%%%%%%%%%%%%%%%%%%%%%%%%%%
\section{Correlated measurements}
%%%%%%%%%%%%%%%%%%%%%%%%%%%%%%%%%%%%%%%%%%%%%%%%%%%%%%%%%%%%%%%%%%
In the case of correlated measurements, the covariance matrix is non diagonal and of the form (\ref{cov}). In its standard estimation, the degree of polarization needs four measurements, however, two images are sufficient to get an estimation of $P^2$ . Indeed, the coefficient $a_1$ is obtained from one measurement, as the coefficient $a_4$ and the squared modulus of $a_2$ can be estimated from the cross-correlation coefficient $\delta_{12}$ between two measurements $I_1$ and $I_2$. We have
\begin{equation}
\delta_{12}= \int \int I_1 I_2 p(I_1,I_2) dI_1 dI_2 = <I_1 I_2>.
\end{equation}
$\delta_{12}$ can be calculated by using the joined density probability function $p(I_1,I_2)$ assuming that ${\bf A}$ is Gaussian circular (i.e. the speckle is supposed to be fully developped).
It can be shown that the correlation coefficient is obtained with
\begin{equation}
\delta_{12}= \frac{1}{det \Gamma c_1^2 c_4^2}  \left( \frac{1 + \frac{|c_2|^{2}}{c_1 c_4}}{\left( 1 - \frac{|c_2|^{2}}{c_1 c_4} \right )^3 }\right )
\end{equation}
where $|c_2|$ stands for the modulus of $c_2$.
Calculating the centered correlation coefficient defined by
\begin{equation}\label{cross_coef0}
\Delta_{12}=  \delta_{12} - <I_1> <I_2> 
\end{equation}
After some simple algebra, we get
\begin{equation}\label{cross_coef}
\Delta_{12}=  |a_2|^2
\end{equation}

Thus the coefficient $|a_2|^2$ can be obtained from two measurements with $ <I_1 I_2>- <I_1> <I_2>$. This remark leads to write the following property.\\

{\bf Property A:} {\it For fully developped speckle fluctuations, the degree of polarization can be obtained from only two intensity images.}
\\\\
One can easily note that the degree of polarization can be written as a function of the OSCI with 
\begin{equation}\label{cor}
P^2(i,j)=  \eta^2(i,j) + 4 \frac{\Delta_{12}}{(<I_1> + <I_2>)^2}
\end{equation}
From (\ref{posci}), the degree of polarization estimated from the OSCI is clearly under-estimated. Thus, we can correct the OSCI in order to get an estimation of the degree of polarization using the correlation coefficient $\Delta_{12}$. \\
In the following part, the different estimation of the degree of polarization are compared to the estimation with four images through simulated data.

%%%%%%%%%%%%%%%%%%%%%%%%%%%%%%%%%%%%%%%%%%%%%%%%%%%%%%%%%%%%%%%%%%
\section{Comparison with numerical experiments}
%%%%%%%%%%%%%%%%%%%%%%%%%%%%%%%%%%%%%%%%%%%%%%%%%%%%%%%%%%%%%%%%%%
We generated $R$ experiments of $N$ samples of complex Jones vectors which follow a Gaussian circular law. The covariance matrix is known and thus $P^2$ is also known. Under the assumption that the statistical average can be estimated by spatial averages in homogenous regions, the coefficients $a_1$  can be estimated from a single image ( $I_1 = |A_X|^2$) like the coefficient $a_4$ ($I_2 = |A_Y|^2$) since
\begin{equation}\label{estima_1}
{\hat a}_1 = \frac{1}{N}\sum ^{N}_{i=1}|A_X(i)|^2
\end{equation}
\begin{equation}\label{estima_2}
{\hat a}_4 = \frac{1}{N}\sum ^{N}_{i=1}|A_Y(i)|^2
\end{equation}
where $\{ A_X(i), A_Y(i) \}$ represents the component of the Jones vector for the sample $i$.\\
The estimation of $P^2$ differs in the studied cases by the way $|a_2|^2$  is estimated. Three different methods are used:
\begin{itemize}

\item Case of four images\\
In this situation, we have both the real and the imaginary part of the coefficient $a_2$. The quantity $|a_2|^2$ is  estimated by

\begin{equation}\label{estim_a2_A}
{\hat \rho}_{A}(i)= \left|\frac{1}{N}\sum ^{N}_{i=1}A_X(i)A^{*}_Y(i) \right|^2 
\end{equation}

\item Case of two images with the OSCI\\
In this case $a_2$ is assumed to be equal to zero.

\item Case of two images with the proposed approach\\
The coefficient  $|a_2|^2$ is estimated using (\ref{cross_coef0}) with

\begin{equation}\label{estim_a2_I}
\begin{array}{lll}
{\hat \rho}_{I}(i) & = & \frac{1}{N}\sum ^{N}_{i=1}|A_X(i)|^2|A_Y(i)|^2 \\
& &\\
&  &- \left( \frac{1}{N}\sum ^{N}_{i=1}|A_X(i)|^2 \right) \left(\frac{1}{N}\sum ^{N}_{i=1}|A_Y(i)|^2\right)
\end{array}
\end{equation}

\end{itemize}

In the three cases, $P^2$ was estimated from the relation (\ref{deg_pol}) with the estimated parameters.\\
In order to characterize the precision of the estimation, one considers six examples of matrix $\Gamma$. 

\begin{equation}
\Gamma_1=\left[ \begin{array}{cc}
15 & 0.2 + 0.5i\\
0.2 - 0.5i & 6
\end{array}\right];  \Gamma_2=\left[ \begin{array}{cc}
16 & 0\\
0 & 3.6
\end{array}\right];
\end{equation}

\begin{equation}
 \Gamma_3=\left[ \begin{array}{cc}
82 & 13i\\
-13i & 17
\end{array}\right];  \Gamma_4=\left[ \begin{array}{cc}
18 & 7 + 8i\\
7 - 8i & 11
\end{array}\right];
\end{equation}

\begin{equation}
\Gamma_5=\left[ \begin{array}{cc}
30 & 16 - 8i\\
16 + 8i & 14
\end{array}\right];  \Gamma_6=\left[ \begin{array}{cc}
1.25 & 5.5i\\
-5.5i & 26
\end{array}\right];
\end{equation}
 These matrices were chosen such as the degree of polarization are approximatively in $\{ 0.2, 0.4, 0.5, 0.6, 0.8, 1 \}$.\\

\begin{figure}
\centerline{\epsfxsize=9cm\epsfbox{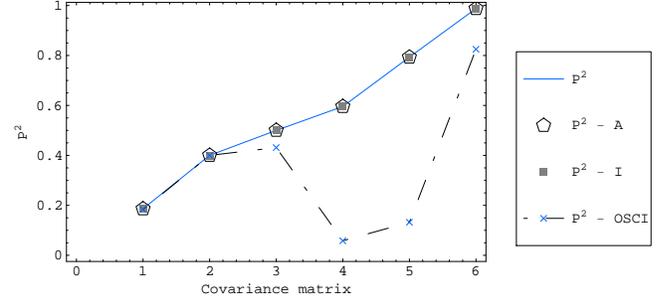}}
\caption{The degree of polarization is plotted as a function of the six covariance matrix for $R = 1000$ and $N = 10000$. $P^2$ is the true degree of polarization, $P^2$ - A is the degree estimated from four measurements, $P^2$ - I is the degree obtained with the numerical simulations using the proposed method for evaluating $|a_2|^2$, $P^2$ - OSCI is the degree estimed from the OSCI.}
\label{fig1}
\end{figure}

\begin{figure}
\centerline{\epsfxsize=9cm\epsfbox{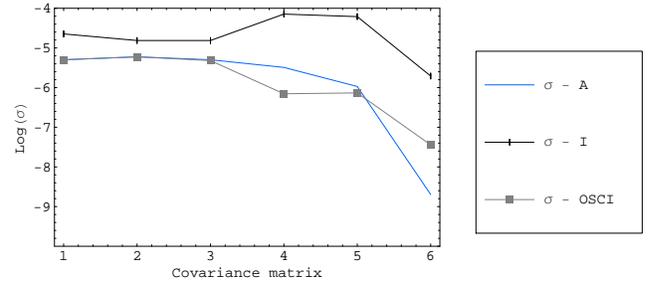}}
\caption{The experimental variance of the degree of polarization is plotted as a function of the six covariance matrix for $R = 1000$ and $N = 10000$. $\sigma$ - A is the standard deviation of the degree of polarization estimated from four measures, $\sigma$ - I is  the standard deviation of $P^2$ obtained with the numerical simulations using the proposed method for evaluating $|a_2|^2$, $\sigma$ - OSCI is the standard deviation of $P^2$ estimed from the OSCI.}
\label{fig2}
\end{figure}

\begin{figure}
\centerline{\epsfxsize=9cm\epsfbox{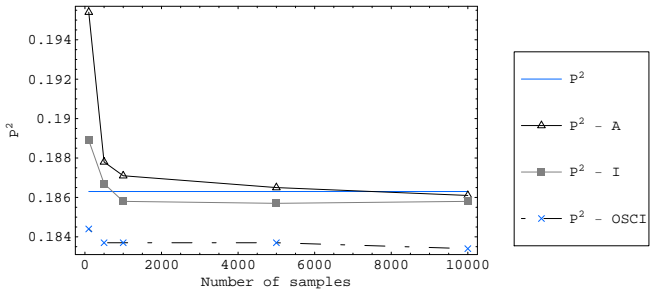}}
\caption{The degree of polarization is plotted as a function of the number of sample $N$ for $R = 1000$ for the covariance matrix $\Gamma_1$. $P^2$ is the true degree of polarization, $P^2$ - A is the degree estimated from four measures, $P^2$ - I is the degree obtained with the numerical simulations using the proposed method for evaluating $|a_2|^2$, $P^2$ - OSCI is the degree estimed from the OSCI.}
\label{fig3}
\end{figure}

\begin{figure}
\centerline{\epsfxsize=9cm\epsfbox{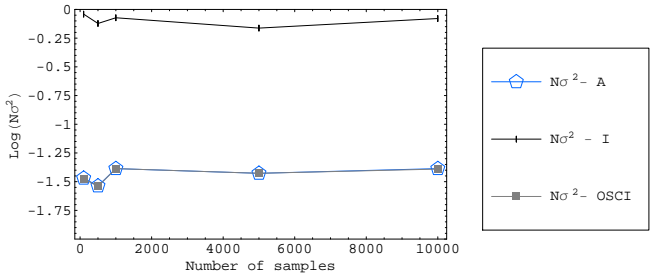}}
\caption{The product of N and the experimental variance of the degree of polarization ($\sigma^2$) is plotted as a function of $N$ for $R = 1000$ for the covariance matrix $\Gamma_1$.  $N \sigma^2$ - A is the product obtained when $P^2$ is estimated from four measures, $N \sigma^2$ - I  is the product obtained when $P^2$ is estimated with the numerical simulations using the proposed method for evaluating $|a_2|^2$, $N \sigma^2$ - OSCI  is the product obtained when $P^2$ is estimated from the OSCI.}
\label{fig4}
\end{figure}

\begin{figure} 
\centerline{\epsfxsize=9cm\epsfbox{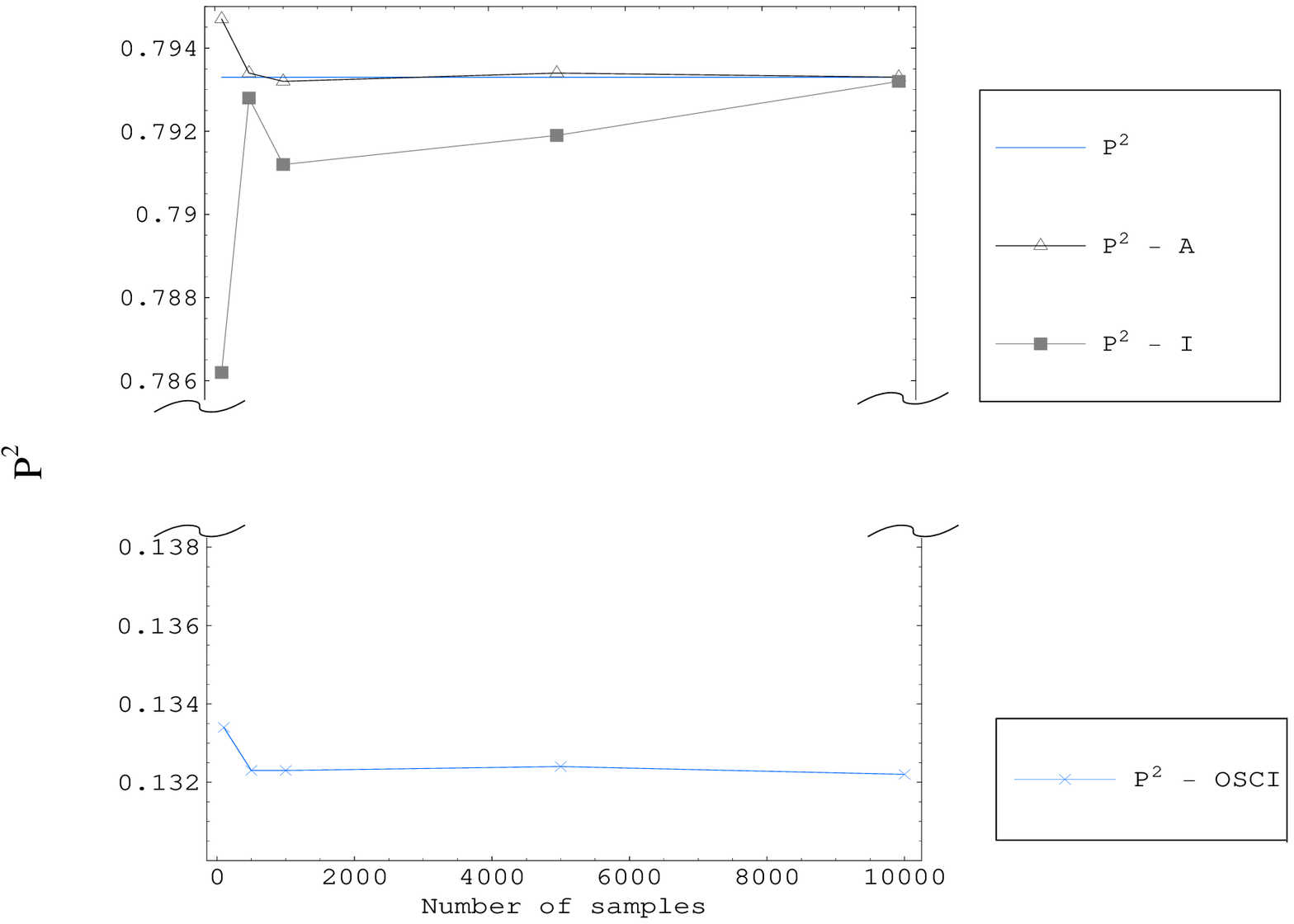}}
\caption{Idem as the Figure 3. using  the covariance matrix $\Gamma_5$ instead of $\Gamma_1$.}
\label{fig5}
\end{figure}

\begin{figure}
\centerline{\epsfxsize=9cm\epsfbox{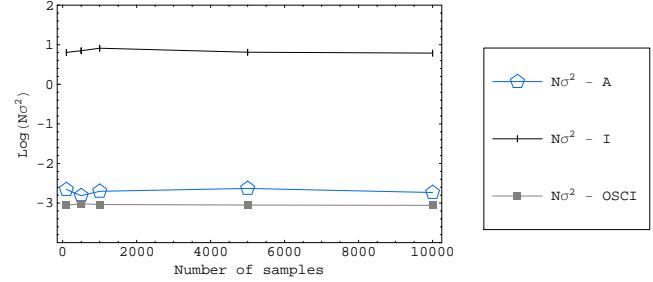}}
\caption{Idem as the Figure 4. using  the covariance matrix $\Gamma_5$ instead of $\Gamma_1$.}
 \label{fig6}
\end{figure}

The simulations are performed for $R = 1000$ realisations of $N = 10000$ samples for the six covariance matrices. For the two matrices $\Gamma_1$ and $\Gamma_5$, supplementary cases have been studied when $N$ $\in$ $\{ 100, 500, 1000, 5000, 10000 \}$. The results are presented in figures \ref{fig1}, \ref{fig2},  \ref{fig3}, \ref{fig4}, \ref{fig5}, \ref{fig6}. Several points are important to notice.
First of all, as expected, the best estimations of the degree of polarization, regarding all the cases tested, was achieved with four images. However the proposed approach that relies on two correlated images produces good estimations of the degree of polarization whatever the covariance matrix is used (fig.\ref{fig1}) as soon as $N > 1000$ (fig.\ref{fig3} and fig.\ref{fig5} ). Note that the estimation with the OSCI gives results which cannot be used if the term $\Delta_{12}/((<I_1> + <I_2>)^2)$ is too high (for example if $|a_2|^2$ is non negligible). Fig.\ref{fig2}, \ref{fig4} and \ref{fig6} show that the experimental variance using four measurements or the OSCI are comparable whatever $\Gamma$ and $N$ are, whereas the variance obtained with the proposed approach is larger. However this precision should be sufficient for some practical applications. This point should be studied in details in a future work.


\section{Conclusion}

We have proposed a new approach to estimate the degree of polarization on polarimetric images degraded by speckle noise. Assuming that the speckle is fully developped, this method allows one to estimate this degree with only two intensity images whereas four images are needed in a standard experimental setup. This presents a great interest in term of reduction of cost of the imagery system since the original setup can be simplified. The proposed approach has been tested on simulated data and compared to the standard estimation techniques that requires either 4 images or 2 independant images (OSCI).  The results show that the proposed method gives good approximation of the degree of polarization.
This study needs to be extended with a theoritical analyses in order to precise the conditions of validity of the proposed approach.  
%%%%%%%%%%%%%%%%%%%%%%%%%%%%%%%%%%%%%%%%%%%%%%%%%%%%%%%%%%%%%%%%%%

%%%%%%%%%%%%%%%%%%%%%%%%%%%%%%%%%%%%%%%%%%%%%%%%%%%%%%%%%%%%%%%%%%

\end{document}